\documentclass[a4paper,fleqn]{cas-dc}

\usepackage[numbers]{natbib}
\usepackage{lineno}
\usepackage{multicol}
\usepackage{graphicx}
\usepackage{color}
\def\tsc#1{\csdef{#1}{\textsc{\lowercase{#1}}\xspace}}
\tsc{WGM}
\tsc{QE}

\begin{document}
\let\WriteBookmarks\relax
\def\floatpagepagefraction{1}
\def\textpagefraction{.001}


\shorttitle{Design of the Global Reconstruction Logic in the Belle~II Level-1 Trigger system}    

\shortauthors{Y.-T.~Lai, T.~Koga, Y.~Iwasaki at al.}  

\title [mode = title]{Design of the Global Reconstruction Logic in the Belle~II Level-1 Trigger system}  

\tnotemark[55665566] 


%

\author[1,2]{Y.-T.~Lai}[orcid=0000-0001-9553-3421]
\cormark[1]
\ead{ytlai@post.kek.jp}

\author[1,2]{T.~Koga}[orcid=0000-0002-1644-2001]
\author[1]{Y.~Iwasaki}[orcid=0000-0001-7261-2557]
\author[3]{Y.~Ahn}[orcid=0000-0001-6820-0576]
\author[1]{H.~Bae}[orcid=0000-0003-1393-8631]
\author[4,5]{M.~Campajola}[orcid=0000-0003-2518-7134]
\author[6]{B.G.~Cheon}[orcid=0000-0002-8803-4429]
\author[6]{H.-E.~Cho}[orcid=0000-0002-7008-3759]
\author[8]{T.~Ferber}[orcid=0000-0002-6849-0427]
\author[8]{I.~Haide}[orcid=0000-0003-0962-6344]
\author[8]{G.~Heine}[orcid=
0009-0009-1827-2008]
\author[9]{C.-L.~Hsu}[orcid=0000-0002-1641-430X]
\author[7]{C.~Kiesling}[orcid=0000-0002-2209-535X]
\author[6]{C.-H.~Kim}
[orcid=0000-0002-5743-7698]
\author[3]{J.~B.~Kim}
[orcid=0000-0002-2072-6082]
\author[3]{K.~Kim}
[orcid=0000-0003-2884-6772]
\author[10]{S.H.~Kim}
[orcid=0000-0002-6860-5781]
\author[10]{I.~S.~Lee}[orcid=0000-0002-7786-323X]
\author[11]{M.~J.~Lee}[orcid=0000-0003-4528-4601]
\author[12]{Y.~P.~Liao}[orcid=0009-0000-1981-0044]
\author[13]{J.~Lin}[orcid=0000-0002-3653-2899]
\author[9]{A.~Little}[orcid=0009-0008-4974-3661]
\author[3]{H.K.~Moon}[orcid=0000-0001-5213-6477]
\author[13]{H.~Nakazawa}[orcid=0000-0003-1684-6628]
\author[8]{M.~Neu}[orcid=0000-0002-4564-8009]
\author[1,2]{S.~Nishida}[orcid=0000-0001-6373-2346] 
\author[8]{L.~Reuter}[orcid=0000-0002-5930-6237]
\author[14]{V.~Savinov}[orcid=0000-0002-9184-2830] 
\author[15]{T.-A.~Sheng}
[orcid=0009-0002-8849-9469] 
\author[13]{J.-G.~Shiu}[orcid=0000-0002-8478-5639]
\author[1]{Y.~Sue}[orcid=0000-0003-2430-8707]
\author[8]{K.~Unger}[orcid=0000-0001-7378-6671]
\author[6]{Y.~Unno}[orcid=0000-0003-3355-765X]
\author[3]{E.~Won}[orcid=0000-0002-4245-7442]
\author[16]{Z.~Xu}[orcid=0009-0005-1048-4744]




\affiliation[1]{organization={High Energy Accelerator Research Organization (KEK)},
            addressline={}, 
            city={Tsukuba},
            postcode={305-0801}, 
            state={Ibaraki}}

\affiliation[2]{organization={SOKENDAI (The Graduate University for Advanced Studies)},
            addressline={}, 
            city={Hayama},
            postcode={240-0193}, 
            state={Kanagawa}}

\affiliation[3]{organization={Korea University},
            addressline={}, 
            city={Seoul},
            postcode={02841}, 
            state={}}

\affiliation[4]{organization={INFN - Sezione di Napoli},
            addressline={}, 
            city={Napoli},
            postcode={I-80126}, 
            state={}}

\affiliation[5]{organization={Universit\`{a} di Napoli Federico II},
            addressline={}, 
            city={Napoli},
            postcode={I-80126}, 
            state={}}

\affiliation[6]{organization={Department of Physics and Institute of Natural Sciences, Hanyang University},
            addressline={}, 
            city={Seoul},
            postcode={04763}, 
            state={Seoul}}

\affiliation[7]{organization={Max-Planck-Institut f\"ur Physik},
            addressline={}, 
            city={M\"unchen},
            postcode={80805}, 
            state={}}
            
\affiliation[8]{organization={Karlsruher Institut f\"ur Technologie~(KIT)},
            addressline={}, 
            city={Karlsruhe},
            postcode={76131}, 
            state={}}

\affiliation[9]{organization={School of Physics, University of Sydney},
            addressline={}, 
            city={Sydney},
            postcode={2006}, 
            state={New South Wales}}

\affiliation[10]{organization={Institute for Basic Science},
            addressline={}, 
            city={Daejeon},
            postcode={34126}, 
            state={}}

\affiliation[11]{organization={Sungkyunkwan University},
            addressline={}, 
            city={Suwon},
            postcode={16419}, 
            state={}}

\affiliation[12]{organization={Institute of High Energy Physics, Chinese Academy of Sciences},
            addressline={}, 
            city={Beijing},
            postcode={100049}, 
            state={}}

\affiliation[13]{organization={Department of Physics, National Taiwan University},
            addressline={}, 
            city={Taipei},
            postcode={10617}, 
            state={Taipei}}

\affiliation[14]{organization={University of Pittsburgh},
            addressline={}, 
            city={Pittsburgh},
            postcode={15620}, 
            state={Pennsylvania}}

\affiliation[15]{organization={Department of Physics, Massachusetts Institute of Technology},
            addressline={}, 
            city={Cambridge},
            postcode={02139}, 
            state={Massachusetts}}

\affiliation[16]{organization={The University of Tokyo},
            addressline={}, 
            city={Tokyo},
            postcode={113-8654}, 
            state={}}

\cortext[1]{Corresponding author}

\fntext[1]{}

\begin{abstract}
The Belle~II experiment is designed to search for physics beyond the Standard Model by investigating rare decays at the SuperKEKB \(e^{+}e^{-}\) collider. 
Owing to the significant beam background at high luminosity, the data acquisition system employs a hardware-based Level-1~Trigger to reduce the readout data throughput by selecting collision events of interest in real time. 
The Belle~II Level-1~Trigger system utilizes FPGAs to reconstruct various detector observables from the raw data for trigger decision-making.
The Global Reconstruction Logic receives these processed observables from four sub-trigger systems and provides a global summary for the final trigger decision. 
Its logic encompasses charged particle tracking, matching between sub-triggers, and the identification of special event topologies associated with low-multiplicity decays. 
This article discusses the hardware devices, FPGA firmware, integration with peripheral systems, and the design and performance of the trigger algorithms implemented within the Global Reconstruction Logic.
\end{abstract}


\maketitle

\section{Introduction}


Belle II~\cite{belle2} at the SuperKEKB accelerator~\cite{superkekb} is a high-luminosity experiment designed to collect a very large data set in the search for physics beyond the Standard Model.
SuperKEKB is an asymmetric-energy collider, operating with 7 GeV electrons and 4 GeV positrons, and is an upgrade of the original KEKB accelerator. 
Based on a nano-beam scheme~\cite{SuperB:2007lel}, SuperKEKB aims to achieve an instantaneous luminosity of $6\times 10^{35}$~cm$^{-2}$s$^{-1}$ and an integrated luminosity of 50~ab$^{-1}$. 

The physics focus of Belle~II includes the study of \(B\) mesons, charm hadrons, \(\tau\) leptons, and the yet hypothetical dark sector~\cite{Belle-II:2018jsg}. 
To efficiently reconstruct their decay products, the Belle~II detector~\cite{belle2} features a general-purpose design with seven sub-detectors. 
Starting from the innermost region at the interaction point~(IP) of SuperKEKB, the Pixel Detector~(PXD), Silicon Vertex Detector~(SVD), and Central Drift Chamber~(CDC) are used to reconstruct the trajectories of charged particles.
The Time-Of-Propagation~(TOP) counter in the barrel region and the Aerogel Ring Imaging Cherenkov~(ARICH) counter in the forward end-cap region are used to identify charged hadrons (protons, kaons, and pions). 
The Electromagnetic Calorimeter~(ECL), consisting of an array of CsI(Tl) crystals, is where photons and electrons deposit most of their energy. 
Surrounding the ECL, a superconducting solenoid generates a magnetic field of 1.5~T. 
The \(K_{L}\) and Muon Detector~(KLM), located in the outermost region of Belle~II, is based on resistive-plate counters and plastic scintillators.
Due to increasing luminosity and higher beam backgrounds from SuperKEKB, the DAQ system is designed to handle a maximum readout throughput of about 3~GB/s~\cite{daq,daq1}. 
To manage this data volume efficiently, Belle~II employs a Level-1 Trigger~(TRG) system~\cite{trg}, which reduces the data throughput by selecting collision events of interest in real time using Field-Programmable Gate Array~(FPGA) devices.

This article describes the Global Reconstruction Logic (GRL), a key component of the Belle~II trigger (TRG) system. 
The GRL performs two primary functions within the TRG system.
\begin{itemize}
    \item First, it receives basic tracking information from the CDC trigger subsystem. 
    This data is then processed and combined to generate CDC-specific trigger bits. 
    \item Second, the GRL integrates the data from all four sub-trigger systems (CDC, ECL, KLM, and TOP) to generate combined sub-detector inputs for the Global Decision Logic (GDL).
\end{itemize}

The paper is organized as follows: Section~\ref{sec:trg} provides an overview of the overall TRG system, including the four sub-trigger systems and the two global decision-making components. 
Section~\ref{sec:grl} details the GRL, including its design, the hardware device, its integration into the TRG system, and the FPGA firmware that supports the trigger algorithms. 
The validation and performance of the trigger algorithm implemented in the GRL are presented in Sec.~\ref{sec:performance}, based on data sets recorded prior to the summer of 2022. 
Finally, Sec.~\ref{sec:summary} summarizes the paper.

\section{Belle~II TRG system}
\label{sec:trg}
\begin{figure*}
    \centering
    \includegraphics[width=0.95\textwidth]{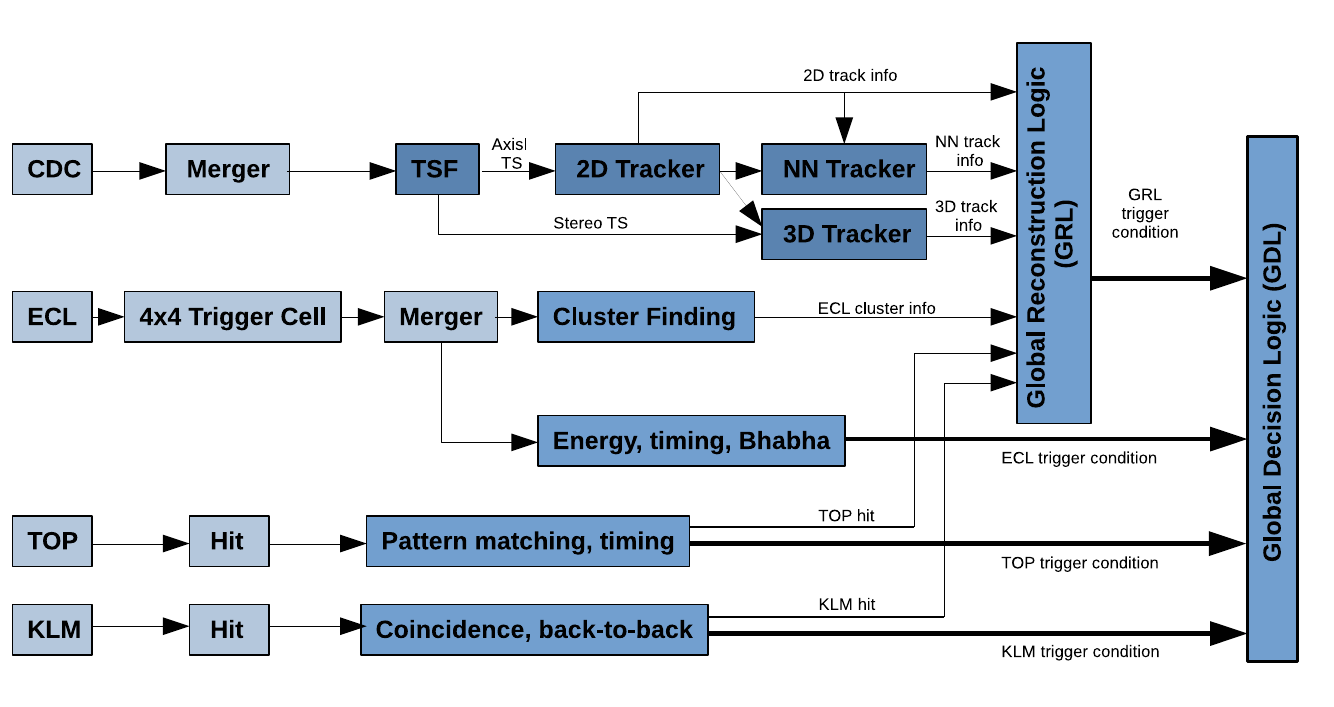}
    \caption{Schematic diagram of the Belle~II Level-1 Trigger~(TRG) system, illustrating its key components and data flow.}
    \label{fg:trg}
\end{figure*}

The Belle II TRG system design is shown in Fig.\ref{fg:trg}. 
The system consists of four sub-trigger components, each utilizing data from a specific detector to observe the signatures of various physics events and accurately measure event timing:
\begin{itemize}
    \item The CDC trigger tracks charged particles to obtain trajectory and kinematic information~\cite{cdctrg, cdctrg1}. 
    \item The ECL trigger measures the deposited energy, event timing, and cluster information from the calorimeter's crystal array~\cite{ecltrg, ecltrg1}, and performs Bhabha scattering identification based on a back-to-back cluster topology. 
    \item The TOP trigger achieves precise event timing by matching the photon arrival time patterns observed on the TOP detector staves with pre-determined patterns from GEANT-4 simulations~\cite{toptrg}.
    \item The KLM trigger uses coincidences and back-to-back correlations among the detector layers to aid in muon identification and provide coarse directional information~\cite{trg}.
\end{itemize}
These four sub-trigger systems receive raw data from the Front-End Electronics (FEE) boards of the respective detectors and transfer their processed outputs to the two-stage global trigger system, consisting of the GRL and GDL, to make the final trigger decision for an event.
All raw data from the Belle II detectors will be read out by the DAQ system according to the timing specified by the TRG system.

\begin{figure}[htb]
\centering
\includegraphics[width=0.45\textwidth]{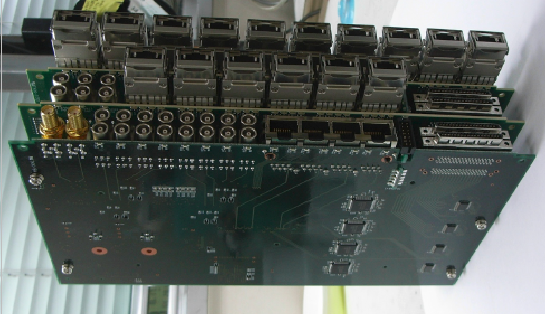}
\caption{Photograph of the Universal Trigger Board~(UT3) used in the Belle~II Level-1 Trigger~(TRG) system.}
\label{fg:ut3}
\end{figure}

The TRG system is composed of FPGA chips, where trigger algorithms are implemented using a Hardware Description Language. 
Except for the Front-End Electronics~(FEE) and Merger devices, all trigger modules use a Universal Trigger Board~(UT3), as shown in Fig.~\ref{fg:ut3}.
The UT3 is based on Xilinx Virtex-6 chips (V6HX380T and V6HX565T)~\cite{virtex6} and features a large number of high-speed serial I/O ports, including 40 GTX~\cite{gtx} and 24 GTH~\cite{gth} ports, as well as a VME bus. 
The entire TRG system operates on a common 127.216~MHz clock (7.8~ns per clock cycle), which is generated by SuperKEKB and distributed through the Trigger and Timing Distribution~(TTD) system~\cite{ttd} of the DAQ. 
Additionally, a common revolution signal, derived from the clock with a period of 10~$\mu$s, is distributed to the TRG through the TTD to ensure synchronization across all modules.

Data is transmitted within the TRG system using FPGA multi-giga serial I/O ports, QSFP modules, and optical fiber cables. To meet the stringent processing latency requirements in the TRG, we developed a user-defined transmission protocol. Our design employs 8B/10B and 64B/66B encoding and supports line rates ranging from 2.5 Gbps to 11.2 Gbps. It features bi-directional transmission for handshake and flow control. The transmission latency is reduced by approximately 50\% compared to the open-source Aurora protocol~\cite{aurora1, aurora2}.
With 8B/10B encoding and a 5.1~Gbps line rate, the latency for our protocol is about 100~ns, while for Aurora it is approximately 200~ns. 
With 64B/66B encoding and an 11.2~Gbps line rate, our protocol achieves a latency of about 100~ns, compared to 300~ns for Aurora. 
This protocol has been implemented across various FPGA chips and their multi-giga transceivers, including Xilinx and Intel FPGAs, within the TRG system, and these components will be explained in more detail later in the paper.
Additionally, the protocol is integrated into the system's flow control and synchronization scheme, as discussed in Section~\ref{sec:cdctrg_control}, further validating the handshake logic design. 
The data transmission quality of the protocol, implemented in UT3, was verified through a Bit Error Rate Test, resulting in a bit error rate of less than $10^{-18}$/s.

The design of the overall TRG system takes the following requirements into account, based on the hardware limitations of both the TRG and the DAQ~\cite{trg}: pipeline processing with no deadtime, an overall latency of less than 4.4~$\mu$s for all processing (including trigger logic and data transmission across all TRG boards), near 100\% trigger efficiency for hadronic events (including $e^{+}e^{-} \to B\bar{B}$ and $e^{+}e^{-} \to q\bar{q}$, where $q = u, d, s, c$), event timing precision of less than 10~ns, event separation of 500~ns, and a limitation on the Level-1 trigger rate of 30~kHz.

In the following, a detailed explanation of the CDC trigger is provided, as the Global Reconstruction Logic~(GRL) serves as the final stage for tracking in the Belle~II Trigger system. The other subsystems (ECL, KLM, and TOP) only provide final trigger objects to the GRL and are therefore described only briefly.

\subsection{CDC Trigger}
The CDC~\cite{cdc} consists of 14,336 sense wires grouped into nine super-layers~(SLs) SL0-SL8 and uses a mixture of helium and ethane as the ionization gas. 
Its alternating sense-wire configuration in each SL enables the reconstruction of particle trajectories in three dimensions.
Each SL comprises 6 to 8 wire layers. 
Super-layers SL0, SL2, SL4, SL6, and SL8 are axial, aligned with the \(z\)-axis (defined as the Belle~II solenoid axis), with SL0 being the innermost layer. 
Super-layers SL1 and SL5 are skewed by approximately 70~mrad, while SL3 and SL7 are skewed in the opposite direction. 
These skewed layers are commonly referred to as stereo layers.


In the CDCTRG system, there are 292 FEE boards, each equipped with a Xilinx Virtex-5 FPGA, and each connected to 48 sense wires~\cite{cdcfee}. 
For each clock cycle of the 127.216~MHz system clock, 1~bit of wire hit information and 3~bits of timing data for each sense wire are read out, with the least-significant bit~(LSB) corresponding to a resolution of approximately 1~ns.
The data flow within the CDCTRG system adopts a 31.8~MHz data clock (31.4 ns for a clock-cycle), which corresponds to four clock cycles of the 127.216~MHz system clock, providing the data rate for refreshing and transferring new data to downstream modules. 
These four system clocks cycles are aligned with the common 10~$\mu$s revolution signal from the TTD, ensuring synchronization across all CDC FEEs. 
The wire hit and timing data within these four clock cycles are merged, with the earliest timing preserved if multiple wire hits occur within the same cycle. 
A 13-bit counter, operating at the 31.8~MHz data clock frequency, provides a time stamp that ranges from 0 to 319. This counter is reset by the common 10~$\mu$s revolution signal from the TTD, maintaining synchronization among all CDC FEE boards.

There are 73 Merger boards based on Intel Arria-II FPGAs that receive and process data from all the CDC Front-End Electronics~(FEEs). The processed data are then transferred to nine Track Segment Finder~(TSF) modules. 
Each TSF corresponds to a super-layer~(SL) of the CDC and identifies wire hits within its SL that match specific patterns, which are then grouped into Track Segments~(TSs). 
These TSs serve as the fundamental elements used by all subsequent tracking modules in the CDCTRG system. 
Following the TSF stage, the trigger electronics for the 2D and 3D trackers are divided into four quadrants, each covering a quarter of the transverse plane of the CDC. 
The four 2D trackers use TS information from the five TSFs corresponding to the five axial SLs, applying a Hough transformation to extract the transverse momentum~(\(p_{t}\)) and incident angle~(\(\phi\)) of full tracks. 
A "full track" is defined as one that passes through all nine SLs of the CDC and reaches the barrel region of the ECL.

The 3D trackers use the corresponding 2D tracker output from each quadrant, along with the Track Segment~(TS) information from the four TSFs of the four stereo SLs. 
The main 3D-track trigger is based on neural networks~(NN), which, for each valid track, provide the longitudinal offset~(\(z_{0}\)) from the interaction point~(IP) and the polar angle~(\(\theta\)), defined with respect to the detector’s solenoid axis~\cite{nn}. 
The neural networks are trained using fully reconstructed data and are implemented in the FPGAs, along with suitable preprocessing of the neural input. 
In parallel, a conventional 3D tracker applies fast linearized track fits on the four stereo track segments' points in the $r$-$z$ plane of the cylindrical coordinate system to obtain $z_{0}$ and $\theta$.
Except for the FEE and Merger boards, all other CDCTRG modules utilize the UT3 board.

\subsection{ECL trigger}
Clustering in the ECLTRG system is based on pattern recognition applied to the hit distributions of groups of crystal cells in the ECL~\cite{ecltrg1}. 
Every 127.2~ns (corresponding to 16 clock cycles of the 127.216~MHz system clock), up to six clusters can be registered, and the information is sent to the Global Reconstruction Logic~(GRL). 
Each cluster contains three pieces of information: a 7-bit \(\theta\) value (ranging from \(0^\circ\) to \(180^\circ\) with a precision of \(1.40625^\circ\)), a 7-bit \(\phi\) value (ranging from \(0^\circ\) to \(360^\circ\) with the same precision), and a 12-bit energy value (with a resolution of 5~MeV).

\subsection{KLM Trigger}  
The barrel part of the KLM detector consists of eight detector slots, forming a regular octagonal transverse cross-section.  
Each slot contains 15 layers.  
The KLM trigger system transmits an 8-bit array representing hit signals from the eight detector slots to the GRL.  
A bit is set when a coincidence is detected in more than six layers within a slot.  
Along with the coincidence information, the GDL also directly receives the back-to-back condition from the KLM trigger.

\subsection{TOP Trigger}  
The TOP detector consists of 16 modules of quartz bars that collect photon hit signals, forming a regular hexadecagonal transverse cross-section.  
The TOPTRG system transmits 16-bit arrays representing hit signals from the 16 detector staves to the Global Reconstruction Logic~(GRL).  
The event timing measured by the TOPTRG system is sent to the GDL.

\section{Firmware Design of the Global Reconstruction Logic}
\label{sec:grl}

\begin{figure*}[htb]
\centering
\includegraphics[width=0.75\textwidth]{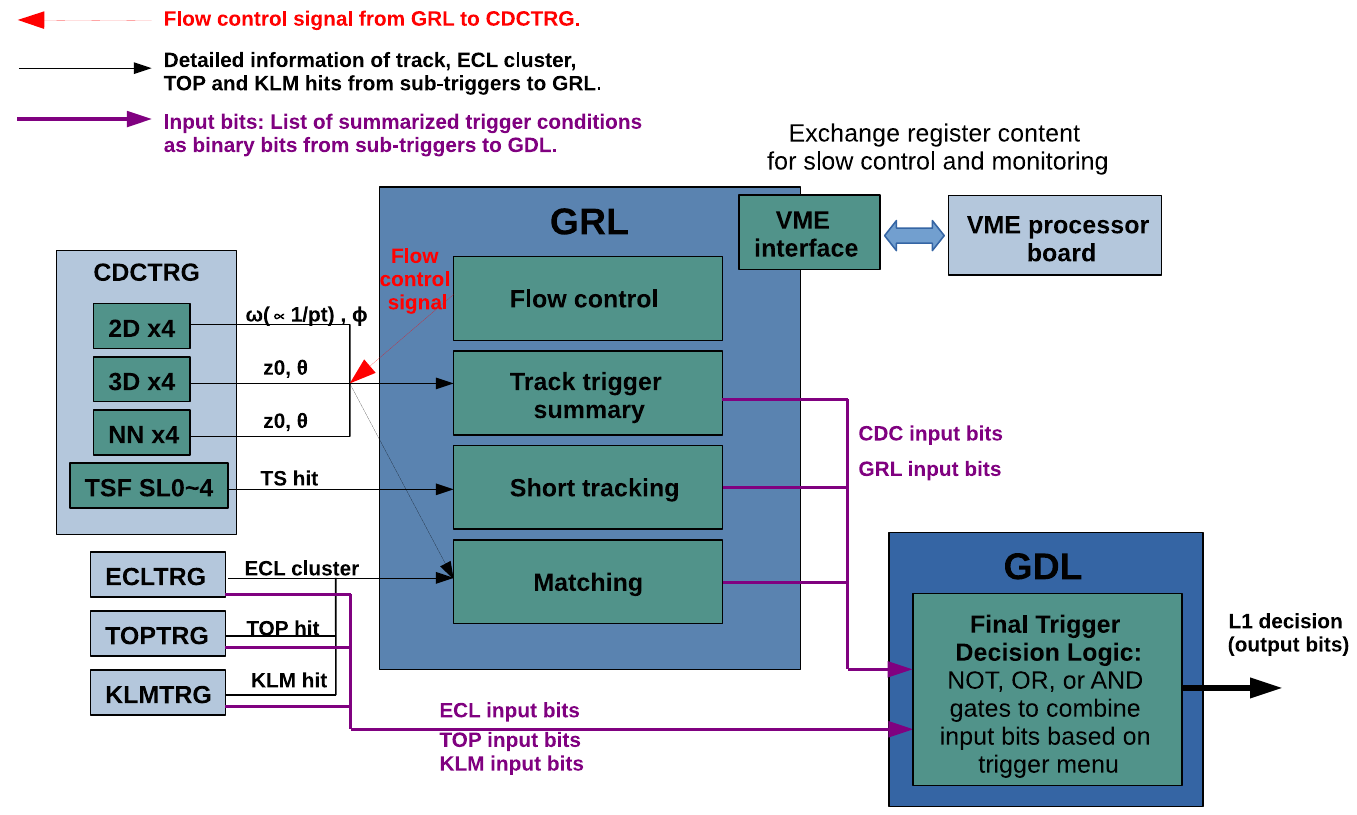}
\caption{Firmware functional block diagrams and data flow of the global trigger system.}
\label{fg:grl}
\end{figure*}

The global trigger system in the Belle~II TRG, comprising the Global Reconstruction Logic~(GRL) and the Global Decision Logic~(GDL), consolidates information from the four sub-trigger systems to make the final trigger decision. 
Since these two modules have distinct functions and responsibilities, they are implemented on separate boards to enhance firmware design flexibility and reduce FPGA resource consumption. 
The GDL primarily receives lists of trigger conditions in the form of binary "input bits," which are combined using logical operations (NOT, OR, and AND) to generate "output bits" representing the Belle~II trigger menu. 
This menu is defined based on the various requirements of Belle~II physics and calibration studies. 
In contrast, the GRL receives detailed trigger observables, such as charged tracks, ECL clusters, and the locations of TOP and KLM hits, from the sub-trigger systems. 
It then extracts and consolidates these trigger conditions before passing them to the GDL. 
Figure~\ref{fg:grl} shows the data flow and firmware functional block diagrams for both the GRL and GDL. 
With input data from separate sub-trigger system modules, the GRL can perform comparisons and matching between them, as well as summarize information across the entire Belle~II acceptance. 
The GRL firmware is implemented on a UT3 board.

The data links of the GRL, which include input data from sub-trigger modules and output data to the GDL, utilize Xilinx Virtex-6 multi-gigabit transceivers on the UT3 boards, QSFP modules, and optical fiber cables, with custom user-defined protocols implemented. 
Table~\ref{tb:data} summarizes the details of each data link in the GRL. 
For certain trigger conditions processed by the GRL before being sent to the GDL, such as CDCTRG track counting and matching between sub-triggers, the TRG's latency limit of 4.4~$\mu$s may be exceeded when these conditions are used in the GDL's final trigger decision. 
To mitigate this, since these trigger conditions are packed into individual bits, LVDS cables are used to transmit them via parallel lines between the GRL and GDL, ensuring they remain within the latency limit.

\begin{table*}[t]
\begin{center}
\caption{Data rate, bandwidth, protocol format, and content of each data link in the GRL. 
The data rate represents the refresh rate of each module based on the 127.216~MHz global system clock. 
Update intervals of 7.8~ns, 31.4~ns, and 125.8~ns correspond to data updates every one, four, and sixteen clock cycles, respectively.}
\begin{tabular}{l|cccccc}
\hline \hline
Module & In/Out & Port & Line rate & Encoding & Data size/rate & content \\
 &  &  & (Gbps) & & (bit)/(ns) &  \\
\hline
CDCTRG 2D ($\times$4) & In & GTH 1 lane ($\times$4) & 5.6 & 64B/66B & 170/31.4 ($\times$4) & $p_{t}$ and $\phi$ of track \\
CDCTRG 3D ($\times$4) & In & GTH 1 lane ($\times$4) & 5.6 & 64B/66B & 170/31.4 ($\times$4) & $z_{0}$ and $\theta$ of track \\
CDCTRG NN ($\times$4) & In & GTH 1 lane ($\times$4) & 5.6 & 64B/66B & 170/31.4 ($\times$4) & $z_{0}$ and $\theta$ of track \\
TSF SL0$\sim$3 ($\times$4) & In & GTX 2 lane ($\times$4) & 5.1 & 8B/10B & 256/31.4 ($\times$4) & TS hit \\
TSF SL4 & In & GTX 3 lane & 5.1 & 8B/10B & 384/31.4 & TS hit \\
ECLTRG & In & GTH 1 lane & 5.6 & 64B/66B & 640/125.8 & position and $E$ of cluster \\
KLMTRG & In & GTH 1 lane & 5.6 & 64B/66B & 42/7.8 & KLM hit \\
TOPTRG & In & GTX 4 lane & 5.1 & 8B/10B & 128/7.8 & TOP hit \\
GDL & Out & GTH 4 lane and LVDS & 5.6 & 64B/66B & 168/7.8 & CDC and GRL input bits \\
\hline \hline
\end{tabular}
\label{tb:data}
\end{center}
\end{table*} 

\subsection{CDCTRG control}
\label{sec:cdctrg_control}
The GRL serves as the central component of the CDCTRG system. 
Within the entire data flow network, the bi-directional nature of the user-defined protocols is leveraged to manage data flow control and synchronization. 
Once the data link of a CDC FEE has been initialized and stabilized, the CDC FEE sends a readiness signal to the corresponding Merger using the protocol. 
Similarly, before reaching the GRL, other CDCTRG modules send a readiness signal to their downstream counterparts once all their data links have been initialized and stabilized, and they have received readiness signals from all their upstream counterparts. 
When the GRL receives readiness signals from all its upstream counterparts, the entire CDCTRG data flow network is considered ready. 
At this point, the GRL waits to receive a revolution signal from the TTD and subsequently propagates a flow control signal to all its upstream modules. Any intermediate module propagates the flow control signal to its upstream modules as soon as it receives it. 
The overall processing latency within the TRG system (from CDC FEE to the GDL's final trigger decision) must be less than the 4.4~$\mu$s limit. Therefore, once all CDC FEEs receive the flow control signals from their downstream modules, they are synchronized to receive the next revolution signal simultaneously. This ensures that all CDC FEEs reset their timestamps and start sending data at the same time. 
The GRL uses a customized VME protocol to read and write register contents via a VME processor board. This functionality allows for monitoring the operation status of the GRL, such as the rate of each input bit, the health of optical links, and data quality. 
It is also used to control the GRL, including resetting optical links, adjusting signal timings, and configuring different algorithms. The exchanged information is integrated into the global slow control system of the Belle~II DAQ~\cite{slc, slc1}.

\subsection{Track Trigger Summary}

\begin{figure*}
    \centering
\includegraphics[width=0.75\textwidth]{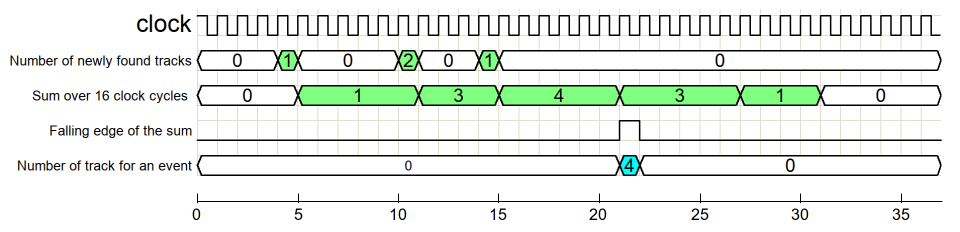}
\caption{Timing diagram of the track counting algorithm implemented in the GRL, showing signal transitions over clock cycles.}
\label{fg:track_counting}
\end{figure*}

For the 2D, 3D, and NN trackers in CDCTRG~\cite{cdctrg, cdctrg1}, each of the four tracker modules covers a quarter of the transverse plane of the CDC. 
For a track identified by a 2D tracker, the tracker calculates the transverse momentum parameter \(\omega\) (which is proportional to \(1/p_{t}\)) and the azimuthal angle \(\phi\). 
The 3D and NN trackers then use additional information to extract the longitudinal impact parameter~(\(z_{0}\)) and the polar angle~(\(\theta\)), respectively. 
The GRL receives track parameters from all twelve modules, which consist of four sets of 2D, 3D, and NN trackers, with each set corresponding to one of the \(\phi\) segments of the CDC. 
In each clock cycle of the 31.8~MHz data clock, each 2D and 3D tracker can identify up to four new tracks (one track per clock cycle per \(\phi\) segment). 
Due to its higher FPGA resource demands, the NN tracker is limited to finding at most one new track per clock cycle.
The average drift time of the CDC is approximately 500~ns. 
Thus, for all tracks corresponding to the same collision event, the wire hit signals and the tracks identified by the tracker modules are contained within a timing window of about 500~ns. 
To count the number of tracks in each event, the GRL employs a track counting algorithm, as illustrated in Figure~\ref{fg:track_counting}. 
In each clock cycle of the 31.8~MHz data clock, the number of newly identified tracks from the 2D trackers is recorded in a shift register with a depth of 16, corresponding to a timing window of approximately 500~ns (\(16\) clock cycles \(\times\) \(31.4\)~ns \(\approx\) \(500\)~ns). 
The GRL also calculates the total number of tracks found over the past 500~ns (i.e., the last 16 clock cycles). 
When a falling edge is detected in the summed track count (i.e., when the track count decreases from the previous clock cycle), the maximum number of tracks for that event is determined and sent to the GDL as part of the input bits.

\begin{figure}[htb]
\centering
\includegraphics[width=0.45\textwidth]{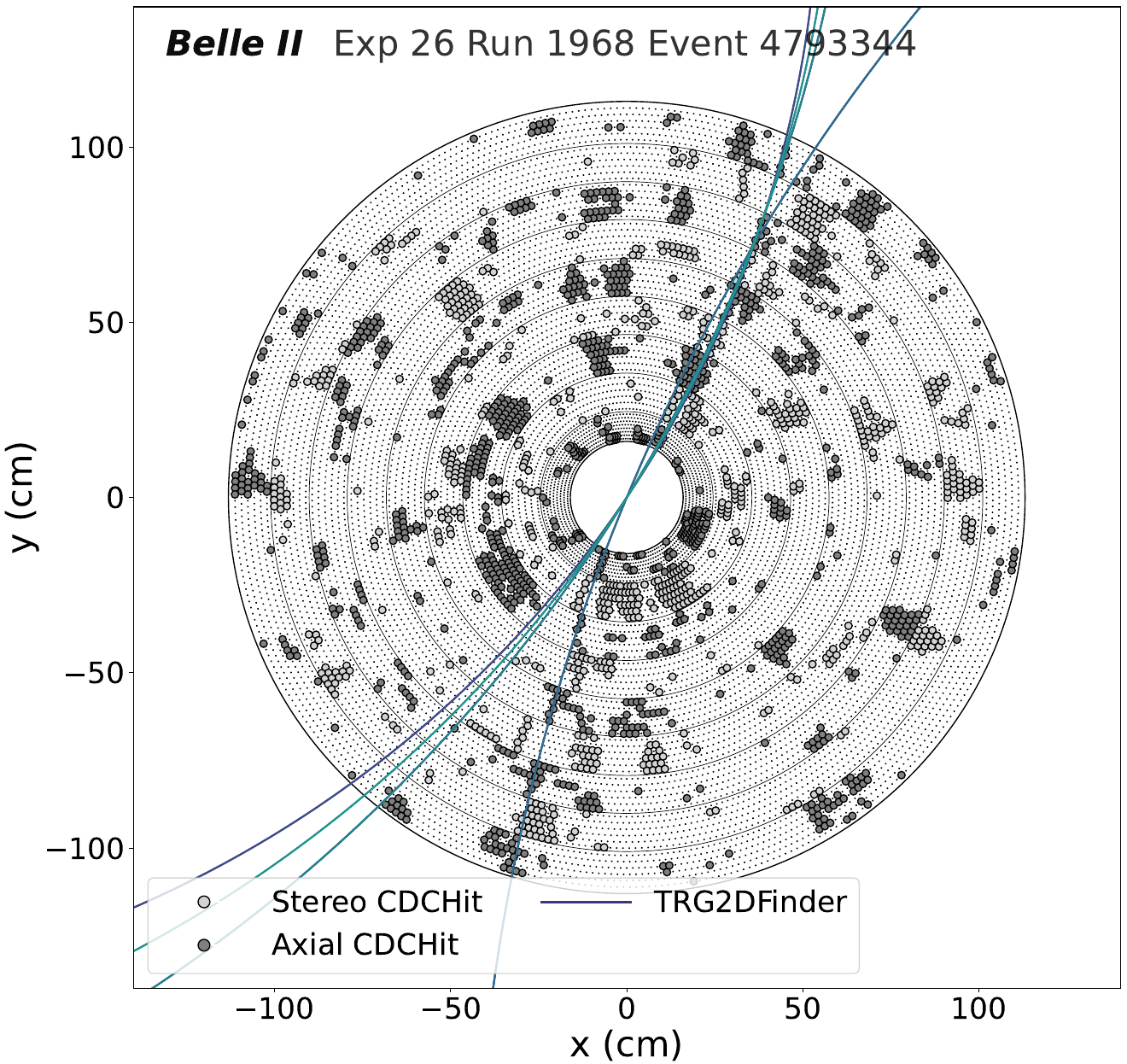}
\caption{Demonstration of cross-talk noise in the CDC in an event display in the $x$-$y$ plane, showing an event from the Belle~II beam collision runs in 2019. The dots indicate the positions of sense wires in the CDC, while the markers represent axial (dark gray) and stereo hits (light gray). The arcs correspond to reconstructed tracks identified by the 2D trackers~(TRG2DFinder) in the CDCTRG system.}
\label{fg:cross_talk}
\end{figure}

The CDC suffers from cross-talk noise in the FEE, which results in multiple noise-induced wire hits occurring simultaneously. This issue causes the 2D trackers to identify numerous duplicate fake tracks, leading to an unexpectedly high track count and negatively impacting track trigger performance. 
Figure~\ref{fg:cross_talk} illustrates this situation, showing multiple wrongly reconstructed tracks due to cross-talk noise.
To mitigate excessive track counting while preserving tracking efficiency, the GRL identifies potential duplicate tracks by comparing the values of \(\omega\) and \(\phi\) between pairs of detected tracks. 
In the 2D trackers~\cite{cdctrg1}, \(\omega\) is defined as \(10.2 / |p_{t}~(\mathrm{GeV})|\) and ranges from \(-33\) to \(33\) in the firmware, while \(\phi\) ranges from \(0\) to \(82\), with each unit corresponding to \(1.125^\circ\). 
If two tracks satisfy \(\Delta\omega < 8\) and \(\Delta\phi < 8\) (corresponding to \(9^\circ\)), they are considered the same track for track counting purposes.

\begin{figure}[htb]
\centering
\includegraphics[width=0.35\textwidth]{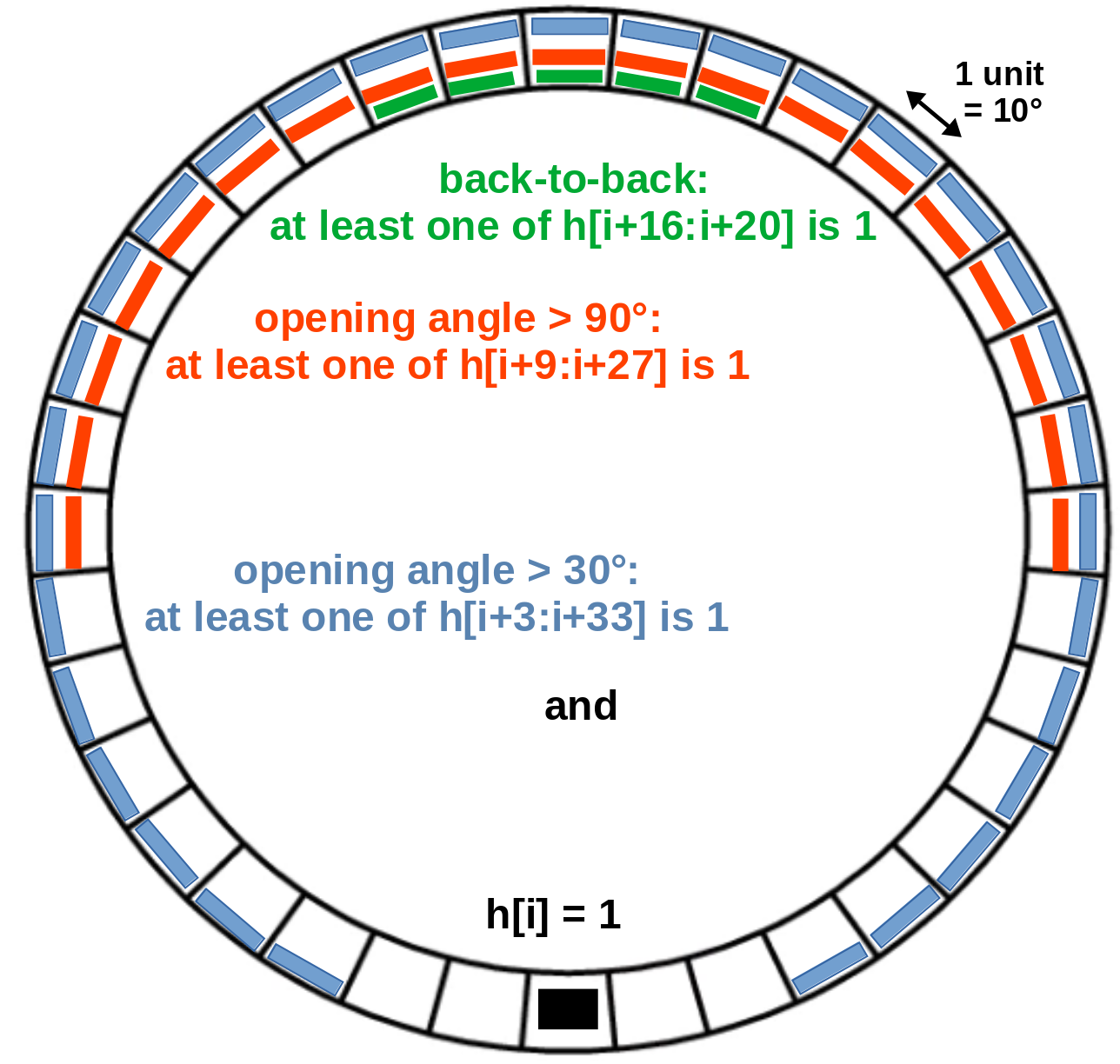}
\caption{Criteria for the three kinds of track geometry conditions: back-to-back, opening angle $>90^{\circ}$, and opening angle $>30^{\circ}$.}
\label{fg:b2b}
\end{figure}

The GRL also provides the geometric relationship between tracks as part of the input bits to the GDL. A 36-bit register is used as a hit array for \(\phi\), where each bit corresponds to a unit of \(10^\circ\). 
When the GRL receives a newly identified track, the corresponding bit in the 36-bit array, based on the \(\phi\) value, is set and remains active for 16 clock cycles. 
This mechanism allows for the observation of coincidences between tracks from the same event. 
Figure~\ref{fg:b2b} illustrates how trigger conditions related to track geometry are determined.

The following track conditions are defined:

\begin{itemize}
    \item \textit{Back-to-back condition}: The \(i^{\text{th}}\) bit is 1, and at least one of the \((i+16)^{\text{th}}\) to \((i+20)^{\text{th}}\) bits is also 1.
    \item \textit{Opening angle \(> 90^\circ\) condition}: The \(i^{\text{th}}\) bit is 1, and at least one of the \((i+9)^{\text{th}}\) to \((i+27)^{\text{th}}\) bits is also 1.
    \item \textit{Opening angle \(> 30^\circ\) condition}: The \(i^{\text{th}}\) bit is 1, and at least one of the \((i+3)^{\text{th}}\) to \((i+33)^{\text{th}}\) bits is also 1.
\end{itemize}

Both the back-to-back and opening angle \(> 90^\circ\) conditions are crucial for triggering a variety of primary physics events, such as hadronic interactions, \(e^+e^- \to e^+e^-\), \(e^+e^- \to \mu^+\mu^-\), \(e^+e^- \to \tau^+\tau^-\), and other processes. 
The opening angle \(> 30^\circ\) condition, in particular, is specifically tailored for triggering a Dark Higgsstrahlung process, in which a dark photon and a dark Higgs boson are produced in electron-positron collisions~\cite{dark}. 
In this scenario, the dark photon remains invisible, while the dark Higgs boson decays into two charged particles with a small opening angle.

Triggers based on track counting in the GRL are crucial for detecting high-multiplicity physics events. 
For example, the three-track trigger and the two-track triggers with an opening angle requirement are key elements for efficiently retaining hadronic events in the TRG system. 
Using a hadronic data sample, the trigger efficiencies for the three-track and two-track opening angle \(> 90^\circ\) conditions are measured to be 90\% and 92\%, respectively. 
When the duplicate track removal scheme is applied to these track conditions, the trigger rates are reduced by approximately 30\% compared to the original ones, with only a 0.5\% decrease in trigger efficiency.

\subsection{Short Track}
\begin{figure}[htb]
\centering
\includegraphics[width=0.25\textwidth]{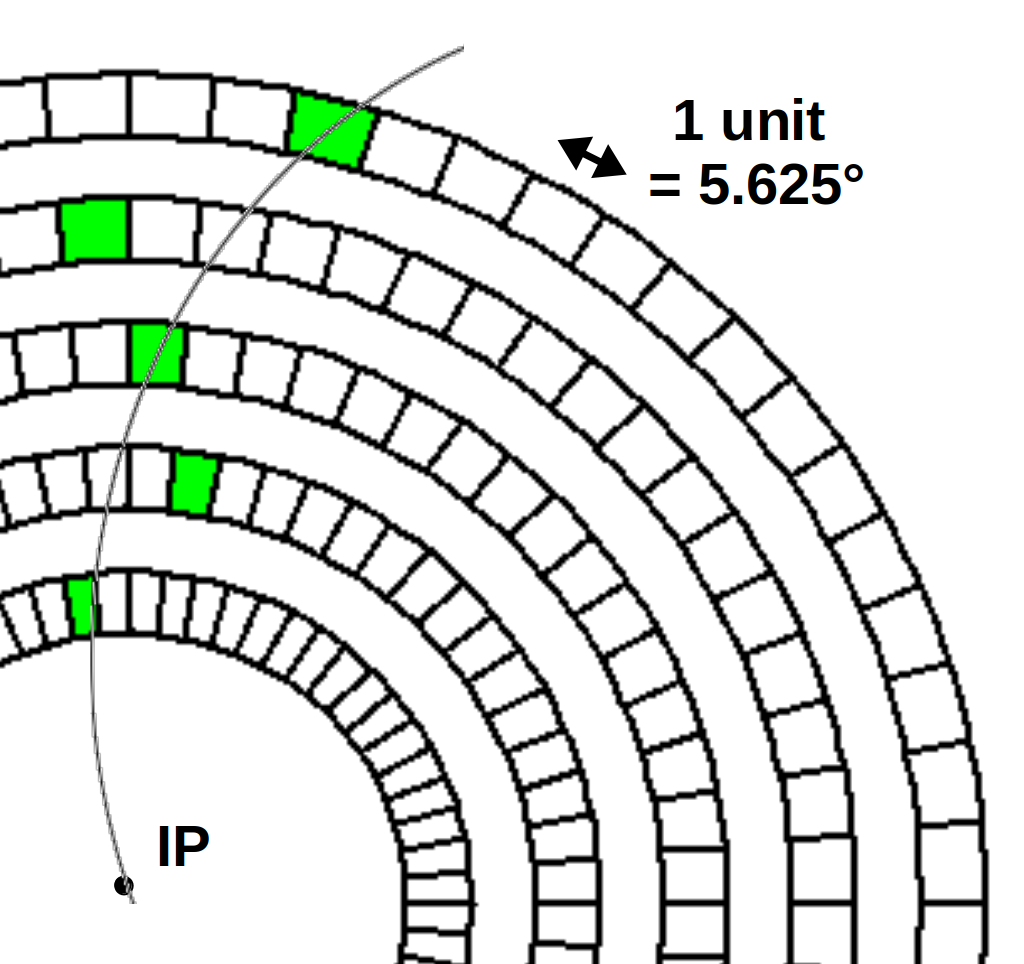}
\caption{Illustration of the short track identification algorithm in the GRL, showing the pattern recognition process using hits across SL0 to SL4, where the green cells represent one example of the 130 possible track patterns used in the algorithm (see text for details).}
\label{fg:st}
\end{figure}

The CDCTRG 2D, 3D, and NN trackers do not reconstruct short tracks that reach the endcap region with fewer than nine SL hits or that curl back inside the CDC~\cite{cdctrg, cdctrg1}. 
To enhance the acceptance of the track trigger toward the endcap region, a short tracking algorithm has been implemented in the GRL firmware using the TS hit information from the five TSF boards of SL0 to SL4. 
The number of TSs across the \(360^\circ\) transverse plane of the CDC is 160, 160, 192, 224, and 256 for SL0, SL1, SL2, SL3, and SL4, respectively.
The greatest common factor among these values, 64, is taken as the mesh size, resulting in a bin width of \(5.625^\circ\). 
This mesh size defines a 2D array of \((5~\text{SL}) \times (64~\text{bits})\) registers to store the TS hits. 
In each clock cycle of the 31.8~MHz data clock, when the GRL receives a newly identified TS from one of the TSFs, it first checks whether the TS has been associated with any full track. 
If it has not, the corresponding bit of the 2D array (based on the TS’s SL and position) is set and remains active for 16 clock cycles. 
The algorithm for identifying short tracks is based on pattern recognition, as illustrated in Figure~\ref{fg:st}. 
To recognize a track pattern, the algorithm first locates a hit in SL0, calculates the distances between the hits in SL1, SL2, SL3, and SL4 relative to the SL0 hit, and defines a pattern using these four integer values. 
For example, Figure~\ref{fg:st} corresponds to the pattern \((3,2,1,4)\). 
A total of 130 possible track patterns have been identified, corresponding to expected hit sequences for tracks originating from the IP.
Once any of these patterns is observed in the 2D array, a new short track is identified, and the \(\phi\) angle corresponding to the SL0 hit is stored. 
The total number of short tracks is then counted. 
Using the \(\phi\) angle, geometric conditions between full tracks and short tracks, as well as between short tracks themselves, are derived and sent to the GDL as part of the input bits.

\subsection{Matching Between Sub-Triggers}
Triggers based on information from a single detector are prone to significant background contamination, leading to a high rate of false triggers. 
For example, many fake tracks in the CDCTRG originate from cross-talk noise in the FEE and beam backgrounds. 
Similarly, in the ECLTRG, the separation between charged and neutral clusters can cause contamination between \(e^+e^- \to e^+e^-\) and \(e^+e^- \to \gamma\gamma\) events.
Since the GRL receives individual observables from each sub-trigger system, it can execute a matching algorithm between hits from outer detectors and tracks from the CDCTRG 2D tracker. 
This matching process helps mitigate background effects.

\begin{figure}[htb]
\centering
\includegraphics[width=0.4\textwidth]{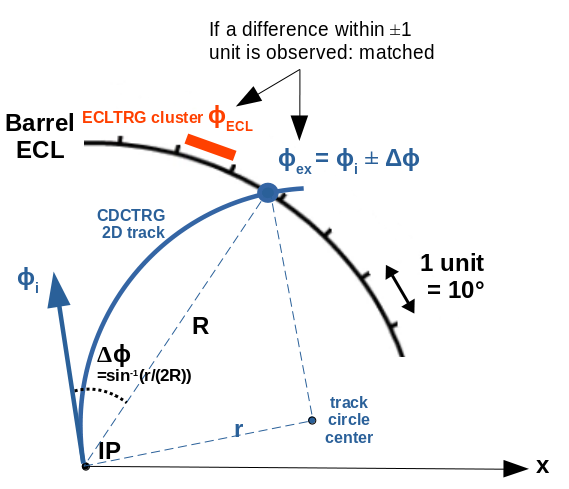}
\caption{Illustration of the matching algorithm between a CDCTRG 2D full track and a ECLTRG cluster in the barrel ECL in the GRL.}
\label{fg:match}
\end{figure}

Figure~\ref{fg:match} illustrates the CDC-ECL matching algorithm using a transverse cut view of the Belle~II detector. 
The parameters $\phi_{i}$ and $\omega$, which are outputs from the 2D trackers to the GRL, are estimated under the assumption that the 2D full track follows a perfect circular arc passing through the interaction point (IP). The radius $r$ of this circular arc is proportional to the transverse momentum $p_{t}$ of the track, where $p_{t} = \frac{10.2}{|\omega|} = 0.0044r$, with $r$ in centimeters and $p_t$ in GeV.
To calculate the angle $\Delta \phi$, the formula $\Delta \phi = \sin^{-1} \left( \frac{r}{2R} \right) = \sin^{-1}(0.0278\omega)$ is used, where $R$ is the fixed distance from the IP to the ECL. 
The resulting azimuthal angle of the 2D full track's extrapolated position at the ECL is given by $\phi_{ex} = \phi_{i} \pm \Delta \phi$, with the sign depending on the charge of the track.

\begin{figure}[htb]
\centering
\includegraphics[width=0.4\textwidth]{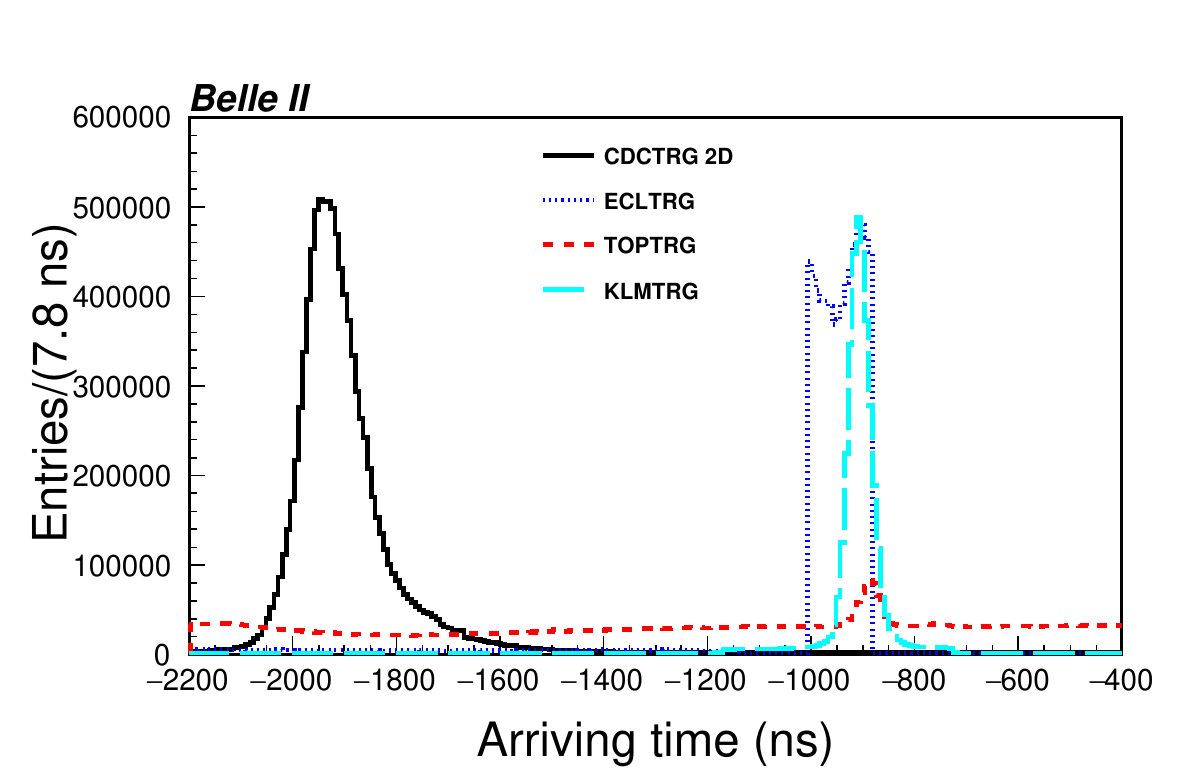}
\caption{Histograms of the arrival times of each sub-trigger module at the GRL relative to the Level-1 L1 trigger.}
\label{fg:arrive}
\end{figure}

In each event, the CDCTRG 2D full track information arrives at the GRL earlier than the information from the other sub-triggers. 
A 36-bit register is allocated as a hit array for \(\phi_{ex}\) of the full tracks, with each unit representing \(10^\circ\). 
When a full track arrives at the GRL, its \(\phi_{ex}\) value is calculated, and the corresponding bit in the 36-bit array is set, persisting for a predefined time period. 
This time period is adjustable to accommodate the latency difference between the CDCTRG and ECLTRG, as demonstrated in Figure~\ref{fg:arrive}.

For clusters within the barrel region \((35^\circ < \theta < 126^\circ)\), the GRL checks whether the corresponding bit and its \(\pm1\) neighboring bits in the 36-bit \(\phi_{ex}\) register are set to determine if a CDCTRG 2D full track matches an ECLTRG cluster. 
This \(\pm1\) criterion, corresponding to a range of three units \((30^\circ)\), is derived from a simulation study using offline software. 
A similar workflow is applied for matching CDCTRG 2D full tracks with TOP and KLM hits, utilizing separate look-up tables for \(\Delta \phi = \sin^{-1}(\frac{r}{2R})\), adjustable \(\phi_{ex}\) persistence time parameters, and matching criteria. 
The persistence time parameters define how long a track’s \(\phi_{ex}\) value remains active in the matching process, ensuring that tracks arriving earlier from CDCTRG remain available long enough to be compared with later-arriving hits from TOP and KLM.

\subsection{Other Trigger Bits from the GRL}
In addition to the trigger bits for major physics processes, this section introduces other trigger bits provided by the GRL for rare physics processes and calibration purposes. 
For instance, track-cluster back-to-back and cluster-cluster back-to-back conditions are implemented using the same criteria shown in Figure~\ref{fg:b2b}. 
These conditions, along with track-cluster matching, are crucial for the systematic study of electron identification. 
Additional triggers require clusters to be in the same hemisphere or the opposite hemisphere as the track, based on their \(\phi\) angle. 
These triggers are used to select \(e^{+}e^{-} \to \tau^{+}\tau^{-}\) events. 
The GRL also searches for coincidences between the innermost three SLs of the CDC (SL0, SL1, and SL2) and KLM hits in the endcap region, using the \(\phi\) angle for matching. 
This trigger line is specifically designed to improve the polar angle acceptance for the muon trigger.

\section{Performance of the GRL trigger logic}
\label{sec:performance}
To validate the performance of the GRL trigger logic, we use various data sets recorded prior to the summer of 2022.  
During real-time data taking with TRG, the data sets are first selected by the standard Level-1 trigger menu, which includes all available trigger conditions defined by the GDL.  
Particles in these data sets are reconstructed using the Belle~II tracking software~\cite{tracking} within the Belle~II offline software framework~\cite{soft,soft1}, where precise track parameters and particle identification information for charged particles are calculated.  
The $e^{+}e^{-} \to \mu^{+}\mu^{-}$ data samples used in this study are selected based on track offset parameters and the muon likelihood parameter for particle identification.  
The absolute efficiency is determined by comparing the number of events with and without a specific trigger condition within a selected data set.

\begin{figure}[htb]
\centering
\includegraphics[width=0.45\textwidth]{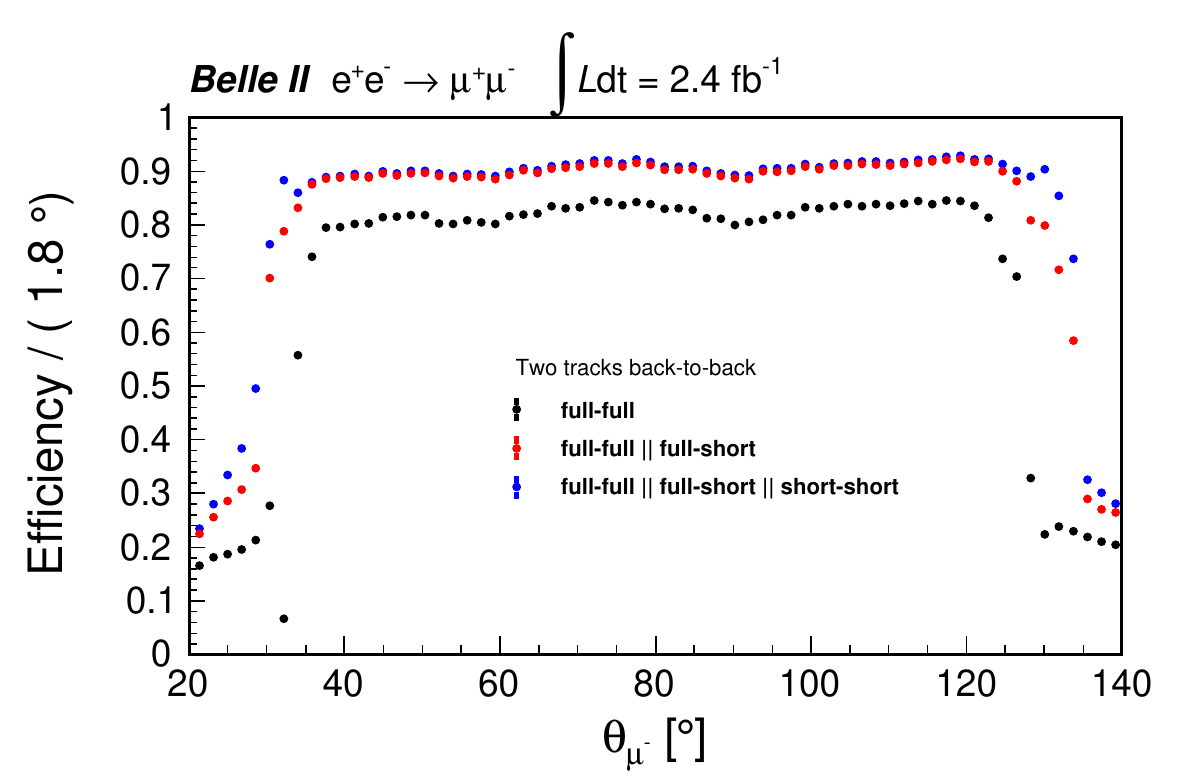}
\caption{Efficiency of different two-track back-to-back triggers as a function of the polar angle \(\theta\) of the \(\mu^{-}\) track. 
The black markers represent cases where both tracks are full tracks. 
The red markers correspond to cases where at most one track can be a short track. 
The blue markers represent cases where both tracks can be short tracks.}
\label{fg:st_theta}
\end{figure}

Using an $e^{+}e^{-} \to \mu^{+}\mu^{-}$ data sample corresponding to 2.4~fb$^{-1}$, the performance of the short tracking algorithm is evaluated. 
Figure~\ref{fg:st_theta} shows the efficiency of different two-track back-to-back triggers, both with and without the contribution from short tracks, as a function of the polar angle of the $\mu^{-}$ ($\theta_{\mu^{-}}$), where the efficiencies of those triggers are estimated within each bin of $\theta_{\mu^{-}}$.
Compared to the full track trigger, an improvement is observed in the endcap region ($\theta_{\mu^{-}}<34^{\circ}$ or $\theta_{\mu^{-}}>127^{\circ}$ in Figure~\ref{fg:st_theta}) due to the additional contribution from short track triggers. 
Furthermore, the inclusion of short tracks improves track finding in the barrel region ($34^{\circ}<\theta_{\mu^{-}}<127^{\circ}$ in Figure~\ref{fg:st_theta}) compared to the full track trigger alone.

\begin{figure}[htb]
\centering
\includegraphics[width=0.45\textwidth]{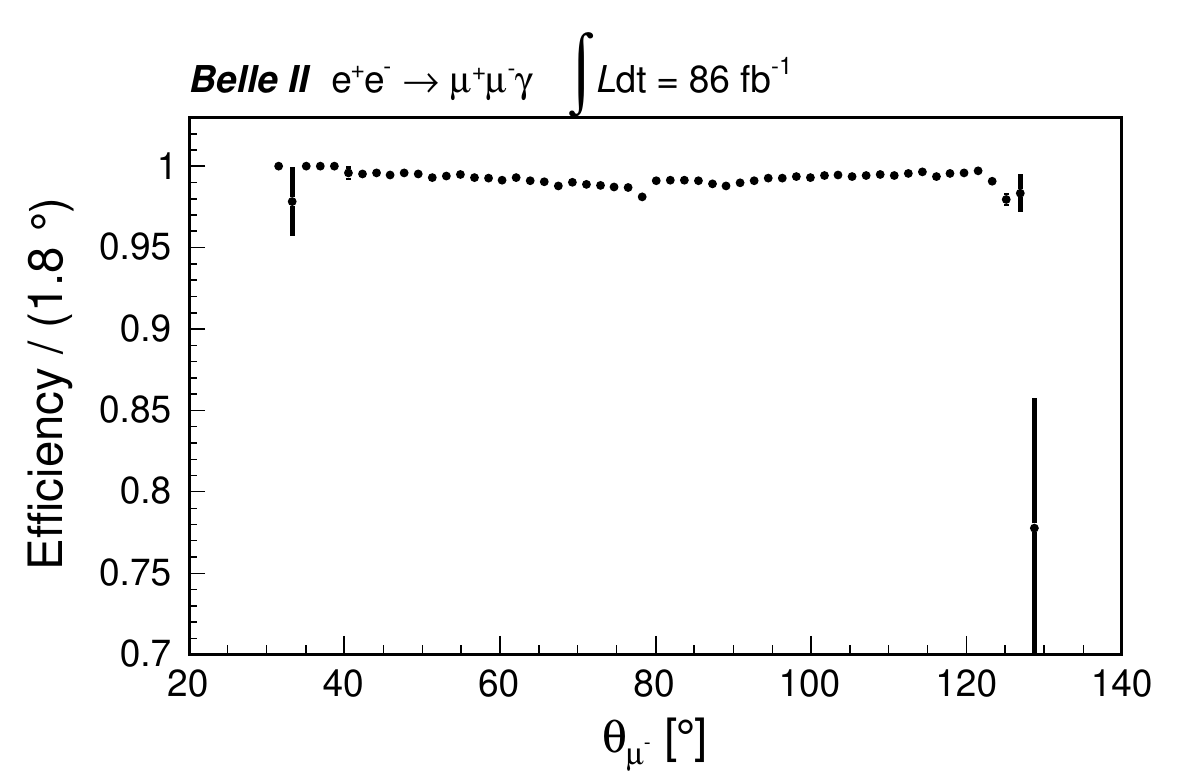}
\caption{Efficiency of the full-track-to-ECL cluster matching trigger as a function of the polar angle \(\theta\) of \(\mu^{-}\).}
\label{fg:cdcecl}
\end{figure}

The performance of full-track-to-ECL cluster matching was verified using an \(e^{+}e^{-} \to \mu^{+}\mu^{-}\gamma\) sample with an integrated luminosity of 86~fb\(^{-1}\). 
Figure~\ref{fg:cdcecl} shows the efficiency as a function of \(\theta\) for \(\mu^{-}\), with an average matching efficiency of 99\%. 

In general, the efficiency for triggering hadronic events is 98\% when using only the CDCTRG conditions and 99\% when using only the ECLTRG conditions. 
When full-track-to-ECL matching is applied in conjunction with the CDCTRG conditions, the efficiency for hadronic events is 97\%. 
Inclusion of full-track-to-ECL cluster matching in the trigger can reduce the trigger rate by a factor of 2 compared to using only CDCTRG conditions.

\begin{figure}[htb]
\centering
\includegraphics[width=0.45\textwidth]{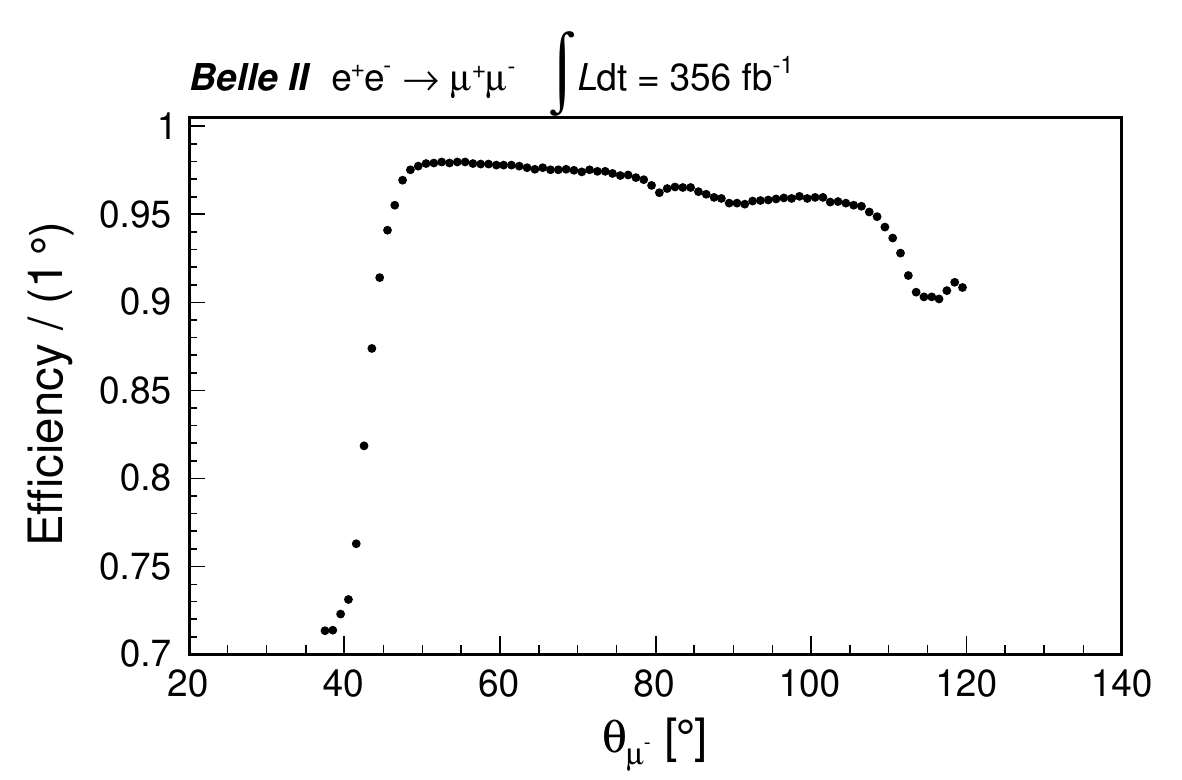}
\caption{Efficiency of the full-track-to-KLM matching trigger as a function of the polar angle \(\theta\) of \(\mu^{-}\).}
\label{fg:cdcklm}
\end{figure}

The performance of full-track-to-KLM matching was verified using an \(e^{+}e^{-} \to \mu^{+}\mu^{-}\) sample with an integrated luminosity of 356~fb\(^{-1}\). 
Figure~\ref{fg:cdcklm} shows the efficiency as a function of \(\theta\) for \(\mu^{-}\), with an average matching efficiency of 94\%. 
Compared to using KLMTRG conditions alone, including full-track-to-KLM matching in the trigger reduces the trigger rate by a factor of 20.
The performance of full-track-to-TOP matching was verified using a hadronic sample, with an average matching efficiency of 92\%. 
Compared to using the TOP trigger alone, including full-track-to-TOP matching in the trigger reduces the trigger rate by a factor of 12.

\section{Summary} 
\label{sec:summary} 
The Belle~II experiment utilizes a hardware-based Trigger~(TRG) system for real-time event selection, reducing the data volume sent to the Data Acquisition~(DAQ) system. 
Within this trigger system, the Global Reconstruction Logic~(GRL) is responsible for consolidating trigger conditions by integrating data from all sub-trigger systems. 
The GRL firmware implements various functionalities, including track reconstruction, matching algorithms, and event topology analysis. 
These conditions are crucial for efficiently identifying a wide range of physics processes, with their performance validated using Belle~II collision data. 
Ongoing hardware upgrades and improvements to the trigger system are expected to enhance the logic design and further optimize the overall performance of the GRL.

\printcredits
This work was supported by National Research Foundation (NRF) of Korea Grants No.~RS-2022-00197659 and No.~RS-2023-00208693, and JSPS KAKENHI Grant Number JP20H05850 and JP20H05858.



\begin{thebibliography}{1}

\bibitem{belle2}
T.~Abe {\it et al.}, (Belle II Collaboration), ``Belle II Technical Design Report,'' arXiv:1011.0352 [physics.ins-det].

\bibitem{Belle-II:2018jsg}
E.~Kou {\it et al.}, ``The Belle II Physics Book,'' {\it PTEP} 2019 12, arXiv:1808.10567 [physics.hep-ex].

\bibitem{superkekb}
K.~Akai, K.~Furukawa, H.~Koiso {\it et al.} (SuperKEKB Collaboration), ``SuperKEKB collider,'' {\it Nucl. Instrum. Methods Phys. Res., Sect. A} vol. 907, pp. 188-199, Nov. 2018.

\bibitem{SuperB:2007lel}
Bona, M. {\it et al.} (SuperB Collaboration), ``SuperB: A High-Luminosity Asymmetric e+ e- Super Flavor Factory. Conceptual Design Report``, {\it SLAC-R-856, INFN-AE-07-02, LAL-07-15, INFN-AE-07-2} 2007 arxiv: 0709.0451  [physics.hep-ex].

\bibitem{daq}
S.~Yamada {\it et al.}, ``Data acquisition system for the Belle II experiment,'' {\it IEEE Trans. Nucl. Sci.}, vol. 62, no. 3, pp. 1175-1180, Jun. 2015.

\bibitem{daq1}
M.~Nakao {\it et al.}, ``Data acquisition system for Belle II,'' {\it J. Instrum.}, vol.5, no. C12004, Dec. 2010.

\bibitem{trg}
Y.~Iwasaki {\it et al.}, ``Level 1 trigger system for the Belle II experiment,'' {\it IEEE Trans. Nucl. Sci.}, vol. 58, no. 4, pp. 1807-1815, Aug. 2011.

\bibitem{cdctrg}
J.-G.~Shiu, ``The level 1 trigger system for Belle II CDC,'' {\it 2016 IEEE-NPSS Real Time Conference (RT)}, Padua, Italy, 2016, pp. 1-3, doi: 10.1109/RTC.2016.7543137.

\bibitem{cdctrg1}
Y.-T.~Lai {\it et al.}, ``Development of the Level-1 track trigger with Central Drift Chamber detector in Belle II experiment and its performance in SuperKEKB 2019 Phase 3 operation,'' {\it J. Instrum.}, vol. 15, no. C06063, Jun. 2020.

\bibitem{ecltrg}
B.G.~Cheon, ``Design of a Electromagnetic Calorimeter Trigger System for the Belle II Experiment,'' {\it JKPS}, Vol. 57, No. 6, pp 1369 - 1375, 2010.

\bibitem{ecltrg1}
S.H.~Kim {\it et al.}, ``Status of the Electromagnetic Calorimeter Trigger system at the Belle II experiment,'' {\it J. Instrum.} vol. 12, no. C09004, Sep. 2017.

\bibitem{toptrg}
L.~Macchiarulo {\it et al.}, ``A probability-optimized fast timing trigger for the Belle II time of propagation detector,'' {\it IEEE Nuclear Science Symposium \& Medical Imaging Conference}, Knoxville, TN, USA, 2010, pp. 630-635.

\bibitem{virtex6}
AMD Xilinx, ``Virtex-6 Family Overview,'' https://docs.amd.com/v/u/en-US/ds150, Aug. 2015.

\bibitem{gtx}
AMD Xilinx, ``Virtex-6 FPGA GTX Transceivers,'' https://docs.amd.com/v/u/en-US/ug366, Jul. 2011.

\bibitem{gth}
AMD Xilinx, ``Virtex-6 FPGA GTH Transceivers,'' https://docs.amd.com/v/u/en-US/ug371, Jun. 2011.

\bibitem{aurora1}
AMD Xilinx, ``Aurora 8B/10B,'' https://docs.amd.com/v/u/en-US/aurora\_8b10b\_ds637, Jan. 2012.

\bibitem{aurora2}
AMD Xilinx, ``Aurora 64B/66B,'' https://docs.amd.com/v/u/en-US/aurora\_64b66b\_ds528, Mar. 2011.

\bibitem{ttd}
M.~Nakao {\it et al.}, ``Timing distribution for the Belle II data acquisition system,'' {\it J. Instrum.}, vol. 7, no. C01028, Jan. 2012.

\bibitem{cdc}
N.~Taniguchi, ``Central Drift Chamber for Belle-II,'' {\it J. Instrum.}, vol. 12, no. C06014, Jun. 2017.

\bibitem{cdcfee}
S.~Shimazaki {\it et al.}, ``Front-end electronics of the Belle II drift chamber,'' {\it Nucl. Instrum. Methods Phys. Res., Sect. A} vol. 735, pp. 193, Sep. 2013.


\bibitem{nn}
S.~B\"ahr {\it et al.}, ``The Neural Network First-Level Hardware Track Trigger of the Belle II Experiment,'' arXiv 2402.14962

\bibitem{dark}
F. Abudin\'en {\it et al.} (Belle II Collaboration), ``Search for a Dark Photon and an Invisible Dark Higgs Boson in $\mu^{+}\mu^{-}$ and Missing Energy Final States with the Belle II Experiment,'' {\it Phys. Rev. Lett.}, 130, 071804, Feb. 2023.

\bibitem{slc}
C.-H. Kim {\it et al.}, ``Trigger slow control system of the Belle II experiment,''  {\it Nucl. Instrum. Methods Phys. Res., Sect. A} vol. 1014, pp. 165748, Oct. 2021.

\bibitem{slc1}
T.~Konno {\it et al.}, ``The Slow Control and Data Quality Monitoring System for the Belle II Experiment,''  {\it IEEE Trans. Nucl. Sci.}, vol. 62, no. 3, Jun. 2015.

\bibitem{tracking}
V.~Bertacchi {\it et al.} (Belle II Tracking Group Collaboration), ``Track finding at Belle II,'' {\it Comput. Phys. Commun.} vol. 259, 107610, Feb. 2021.

\bibitem{soft}
T~.Kuhr {\it et al.}, ``Belle II Software,'' {\it J. Phys.: Conf. Ser.} {\textbf 762} 012007, 2016.

\bibitem{soft1}
T~.Kuhr {\it et al.}, ``The Belle II Core Software,''  {\it Comput Softw Big Sci} {\textbf 3} 1, 2019. 

\end{thebibliography}


\appendix

\end{document}